\documentclass[aps,amsmath,amssymb,floatfix,nofootinbib]{revtex4}
\usepackage{graphicx}
\usepackage{amssymb} %%%%%%%%%%%%%%%%%%%%%%%%%%%%%%%%%%%%%%%%%%%%%%%%%%%%
\begin{document}
\title{Reaction mechanisms in transport theories: a test of the nuclear
effective interaction}

\author{M Colonna${}^{1}$, V Baran${}^{2}$,
M Di Toro${}^{1,3}$, B Frecus${}^{2}$, YX Zhang${}^{4}$}

\address{${}^{1}$Laboratori Nazionali del Sud, INFN, via Santa Sofia 62, 
I-95123, Catania, Italy}
\address{${}^{2}$Physics Faculty, University of Bucharest, Romania}
\address{${}^{3}$Physics-Astronomy Dept., University of Catania, Italy}
\address{${}^{4}$China Institute of Atomic Energy, P.O. Box 275 (10), Beijing 102413, P.R. China}

%\ead{colonna@lns.infn.it}

\begin{abstract}
We review recent results concerning collective excitations in neutron-rich 
systems and reactions between charge asymmetric systems at Fermi
energies. %ranging from 30 MeV/u to 1 GeV/u.
%A very important open question related to the pygmy dipole resonance is about
%its quite elusive collective nature. In this paper, 

Solving numerically self-consistent transport equations for neutrons and
protons with specific initial conditions, we explore the structure
of the different dipole vibrations in the $^{132}Sn$ system and
investigate their dependence on the symmetry energy. We evidence the
existence of a distinctive collective mode, that can be associated with
the Pygmy Dipole Resonance, with an
energy well below the standard Giant Dipole Resonance and isoscalar-like
character, i.e. very weakly dependent
on the isovector part of the nuclear effective interaction. At
variance, the corresponding strength is rather sensitive to the
behavior of the symmetry energy below saturation, which rules the
number of excess neutrons in the nuclear surface.

In reactions between charge asymmetric systems at Fermi
energies,  %ranging from 30 MeV/u to 1 GeV/u.
we investigate the interplay between dissipation mechanisms
and isospin  effects. 
Observables sensitive to the isospin dependent part of nuclear interaction
are discussed, providing information on the symmetry energy density dependence
below saturation.   

\end{abstract}

\maketitle

\vskip -1.0cm
\section{Introduction}
%: The Elusive Symmetry Term of the EOS}
The Equation of State (EOS) of nuclear matter 
%The behavior of nuclear matter under several conditions of density and temperature
is of crucial importance for the understanding of a large variety of phenomena
in nuclear physics and astrophysics.    
%ranging from the structure of nuclei and their decay modes, up to the life and the
%properties of massive stars.
%The EOS of nuclear matter plays a fundamental role
%in the understanding of many aspects of nuclear physics and astrophysics. 
Transient states of nuclear matter far from normal conditions can be created
in terrestrial laboratories and 
many experimental and theoretical efforts have been devoted 
to the study of nuclear reactions, from low to intermediate energies, as a 
possible tool to learn about the behavior of nuclear matter and its EOS.
In particular, the availability of exotic beams has opened
the way to explore, in laboratory conditions, new aspects of nuclear structure
and dynamics up to extreme ratios of neutron (N) to proton (Z) numbers.
Over the past years, 
measurements of isoscalar collective vibrations, collective flows and
meson production have contributed to constrain the EOS for symmetric matter
for densities up to five time the saturation value \cite{cons}. However, the EOS of
asymmetric matter has comparatively few experimental constraints: The
isovector part of the nuclear effective interaction (Asy-EOS) 
and the corresponding
symmetry energy are largely unknown 
far from normal density. 
%We recall that 
%the symmetry energy $E_{sym}$ appears in the energy density
%$\epsilon(\rho,\rho_3) \equiv \epsilon(\rho)+\rho E_{sym} (\rho_3/\rho)^2
% + O(\rho_3/\rho)^4 +..$, expressed in terms of total ($\rho=\rho_p+\rho_n$)
% and isospin ($\rho_3=\rho_p-\rho_n$) densities. The symmetry term gets a
%kinetic contribution directly from basic Pauli correlations and a potential
%part from the highly controversial isospin dependence of the effective 
%interactions.

This information is essential 
also in the astrophysical context, for the understanding of the properties of
compact objects such as neutron stars, whose crust behaves as low-density 
asymmetric nuclear matter \cite{Lattimer} and whose core may touch extreme
values of density and asymmetry. 
Moreover, the low-density behavior of the symmetry energy also affects
the structure of exotic nuclei and 
%features such as the appearance of 
%new collective modes involving the neutron skin. 
the appearance of new features involving the neutron skin \cite{Colo}.

Over the past years,  several observables which
are sensitive to the Asy-EOS and testable
experimentally, have been suggested
\cite{Isospin01,baranPR,WCI_betty,baoPR08}.
%We remind that the knowledge of the 
%EOS of asymmetric matter is very important at low densities (neutron skins,%
%nuclear structure at the drip lines, neutron distillation in fragmentation,
% neutron star formation and crust..) as well as at high densities (transition 
%to a deconfined phase, neutron star mass/radius, cooling, hybrid structure,
% formation of black holes...). 

%Nuclear reactions from low to Fermi energies
% will bring information on the symmetry term around (below) normal density, 
%while intermediate energies will probe high density regions.

In this contribution we will focus on some effects related to the low-density sector
of the nuclear EOS. The first part will be devoted to the study of collective
excitations in neutron-rich systems. New exotic collective modes show up when one 
moves away from the valley of stability \cite{ref1}. In particular, we will discuss results 
concerning the appearance of 
an interesting exotic mode, the Pygmy Dipole Resonance (PDR), which was observed as
an unusually large concentration of the dipole response at energies
clearly below the values associated with the standard Giant Dipole Resonance (GDR).
Then, in the second part, we will discuss dissipation and
fragmentation mechanisms at Fermi energies, concentrating on isospin effects
and on the sensitivity of detectable observables to the Asy-EOS.

%The construction of an $Hadron-EOS$ at high baryon and 
%isospin densities will finally allow the possibility of developing a model 
%of a 
%hadron-deconfinement transition at high density for an asymmetric matter
%\cite{ditoro_dec}. The problem of a correct treatment of the isospin in a
%effective partonic $E0S$ will be stressed.  

\section{Transport theories and symmetry energy}

%at densities below and around the saturation 
%value. This regime of densities is important for studies of the structure
%of exotic nuclei, of the neutron star crust, and for supernova explosions, where
%a key issue is the clustering of low-density matter. 
%On the other hand, reactions between charge asymmetric systems at relativistic energies 
%allow one to create transient
%states of nuclear matter at high density, temperature and asymmetry and to access
%the high density behavior of the symmetry energy. 

Nuclear collective motion and
nuclear reactions are modeled by solving transport equations
%, that describe
%the evolution of the one-body density in response to the action of a 
%mean-field potential and to the effects of the residual two body interaction.
%The transport codes are 
based on 
mean field theories, with correlations included via hard nucleon-nucleon
elastic collisions and via stochastic forces, selfconsistently
evaluated from the mean phase-space trajectory, see 
\cite{baranPR,chomazPR}. 
Stochasticity is 
essential in 
order to get distributions as well as to allow for the growth of dynamical 
instabilities. 

In the beam energy range up to a few hundred $MeV/u$, the appropriate tool is the 
so-called Boltzmann-Langevin equation (BLE) \cite{chomazPR}:
\begin{equation}
{{df}\over{dt}} = {{\partial f}\over{\partial t}} + \{f,H\} = I_{coll}[f] 
+ \delta I[f],
\end{equation}
 where $f({\bf r},{\bf p},t)$ is the one-body distribution function, 
the semi-classical analog of the Wigner transform of the one-body density, 
$H({\bf r},{\bf p},t)$ the mean field Hamiltonian, 
$I_{coll}$ the two-body collision term 
incorporating the Fermi statistics of the particles,
and 
$\delta I[f]$ the fluctuating part of the
collision integral. %\cite{Ayik,Randrup}.
Here we follow the approximate treatment to the BLE introduced in \cite{SMF}, the so-called
Stochastic Mean Field (SMF) model, where fluctuations are injected just in coordinate
space by agitating the density profile. 

%We recall that 
The symmetry energy, $E_{sym}$, appears in the energy density
$\epsilon(\rho,\rho_i) \equiv \epsilon(\rho)+\rho E_{sym}/A~ (\rho_i/\rho)^2
 + O(\rho_i/\rho)^4 +..$, expressed in terms of total ($\rho=\rho_p+\rho_n$)
and isospin ($\rho_i=\rho_p-\rho_n$) densities.
$E_{sym}$ gets a
kinetic contribution directly from basic Pauli correlations and a potential
part, $C(\rho)$,  from the highly controversial isospin dependence of 
the effective interactions: 
\begin{equation}
\frac{E_{sym}}{A}=\frac{E_{sym}}{A}(kin)+\frac{E_{sym}}{A}(pot)\equiv 
\frac{\epsilon_F}{3} + \frac{C(\rho)}{2\rho_0}\rho .
\end{equation}
The sensitivity of the simulation results can be tested against 
different choices of the density dependence of the
coefficient $C(\rho)$.
%isovector part of the EOS. 

\vskip -1.0cm

\section{Collective excitations in neutron-rich systems}

 One of the important tasks in many-body physics is to understand the emergence of the
collective features as well as their structure in terms of the individual motion of the
constituents.  The experimental characterization and theoretical
description of new exotic collective excitations
is a challenge for modern nuclear physics. Recent experiments provided
several evidences  about their existence but the available information is still
incomplete and their nature is a matter of debate.
In particular, many efforts have been devoted to the study of the PDR, 
identified as
an unusually large concentration of the dipole response at energies
below the values corresponding to the 
GDR. The latter is one of the most prominent and robust collective
motions, present in all nuclei, whose centroid position varies, for
medium-heavy nuclei, as $80 A^{-1/3} MeV$.
 From a comparison of the
available data for stable and unstable $Sn$ isotopes a correlation
between the fraction of pygmy strength and isospin asymmetry was
noted \cite{ref2}. In general the exhausted sum-rule
increases with the proton-to-neutron asymmetry. This behavior was
related to the symmetry energy properties below saturation and
therefore connected to the size of the neutron skin
\cite{ref3,ref4,Colo}.

 In spite of the theoretical progress in the interpretation of
this mode and new experimental
information \cite{ref6,ref7,ref8,ref9}, a number of critical
questions concerning the nature of the PDR still remains.
Our goal is to address the important issue related
to the collective nature of the PDR in connection with the role of the symmetry energy.

An accurate picture of the GDR in nuclei corresponds to an admixture of
Goldhaber-Teller (G-T) and Stenweidel-Jensen (S-J) vibrations. The latter, in
symmetric nuclear matter, is a volume type oscillation of the
isovector density $\rho_i= \rho_n - \rho_p$  keeping the total
density  $\rho=\rho_n+\rho_p$ constant \cite{ref10}.
% Then it is governed just by the isovector part of the nuclear interaction.
A microscopic, self-consistent study of the collective features
and of the role of the nuclear effective interaction
upon the PDR can be performed within the Landau theory of Fermi liquids.
This is based on two coupled Landau-Vlasov kinetic equations (see Eq.(1), neglecting
the stochastic term) for neutron and proton one-body
distribution functions $f_q(\vec{r},\vec{p},t)$ with $q=n,p$, %:
%\begin{equation}
%\frac{\partial f_q}{\partial t}+\frac{\bf p}{m}\frac{\partial f_q}{\partial {\bf r}}-
%\frac{\partial U_q}{\partial {\bf r}}\frac{\partial f_q}{\partial {\bf p}}=I_{coll}[f] ,
%\label{vlasov}
%\end{equation}
and was applied  quite successfully in describing various features of the GDR,
including pre-equilibrium dipole excitation in fusion reactions \cite{ref11}.
%\cite{baran2001},
%or in the description of transition from isovector zero sound to first sound  mode
%in nuclear matter \cite{baran1999}.
However, it should be noticed that within such a semi-classical description
shell effects are absent, certainly important in shaping the fine structure
of the dipole response \cite{ref12}.
By solving numerically the Vlasov equation in the absence of Coulomb interaction, 
Urban \cite{ref13} evidenced from the study of the total dipole moment $D$ a collective response 
around $8.6$ $MeV$ which was identified as a pygmy mode. It was pointed out, from the properties of
transition densities and velocities, that the PDR can be related to one of the low-lying modes 
associated with isoscalar toroidal excitations, providing  indications about its isoscalar character.
Here, considering in the transport simulations also the Coulomb interaction, we
shall investigate in a complementary way the collective nature of PDR by studying the 
dynamics of the pygmy degree of freedom, $D_y$, that is usually associated with the neutron excess in the nuclear surface \cite{ref17}. 
%suggested by HOSM. 
Moreover, we shall explore the
isoscalar character of the
mode by a comparative analysis employing three different density parametrizations of the 
symmetry energy.

\subsection{Ingredients of the simulations}
We neglect the two-body collision effects and hence the main ingredient of
the dynamics is the nuclear mean-field, for which we consider a Skyrme-like ($SKM^*$) parametrization
$\displaystyle U_{q} = A\frac{\rho}{\rho_0}+B(\frac{\rho}{\rho_0})^{\alpha+1} + C(\rho)
\frac{\rho_n-\rho_p}{\rho_0}\tau_q
+\frac{1}{2} \frac{\partial C}{\partial \rho} \frac{(\rho_n-\rho_p)^2}{\rho_0}$,
%\begin{align}
%U_{q}&=A\frac{\rho}{\rho_0}+B(\frac{\rho}{\rho_0})^{\alpha+1} + C(\rho)
%\frac{\rho_n-\rho_p}{\rho_0}\tau_q + \nonumber  \\
%&+\frac{1}{2} \frac{\partial C}{\partial \rho} \frac{(\rho_n-\rho_p)^2}{\rho_0},
%\label{meanfield}
%\end{align}
where $\tau_q = +1 (-1)$ for $q=n (p)$ and $\rho_0$ denotes the saturation density.
The saturation properties of symmetric nuclear matter are reproduced
with the values of the coefficients
%appearing in Eq. (\ref{meanfield})
$A=-356.8 MeV$, $B=303.9 MeV$, $\alpha=1/6$,
leading to a compressibility modulus $K=201 MeV$. For the isovector sector we employ
three different parameterizations of $C(\rho)$ with the density: the asysoft,
the asystiff and asysuperstiff respectively, see \cite{baranPR} for a detailed description.
The value of the symmetry energy,
$\displaystyle E_{sym}/A $, %= {\epsilon_F \over 3}+{C(\rho) \over 2}{\rho \over \rho_0}$,
at saturation, as well as
the slope parameter, $\displaystyle L = 3 \rho_0 \frac{d E_{sym}/A}{d \rho} |_{\rho=\rho_0}$,
are reported in Table \ref{table1} for each of these Asy-EOS. Just below the saturation density
the asysoft mean field has a weak variation with density while the asysuperstiff shows
a rapid decrease. 
Then, due to surface
contributions to the collective oscillations, we expect to see
some differences in the energy position of the dipole response of the system.

The numerical procedure to integrate the transport equations is based on the
test-particle (t.p.) method. For a good spanning of phase-space we work with $1200$ t.p. per nucleon.
We consider the
neutron rich nucleus $^{132}Sn$ and we
determine its ground state configuration as the equilibrium (static)
solution of Eq.(1). 
%(\ref{vlasov}). 
Then  proton and neutron densities
%\begin{equation}
$\displaystyle \rho_q(\vec{r},t)=\int \frac{2 d^3 {\bf p}}{(2\pi\hbar)^3}f_q(\vec{r},\vec{p},t)$
%\label{dens}
%\end{equation}
can be evaluated.
As an additional check of our initialization procedure,
the neutron and proton mean square radii
%$\sqrt{\langle r_q^2 \rangle}$,
%\begin{equation}
$\displaystyle \langle r_q^2 \rangle = \frac{1}{N_q} \int r^2 \rho_q(\vec{r},t) d^3 {\bf r}$,
%\label{rmspn}
%\end{equation}
as well as the skin thickness
$\displaystyle \Delta R_{np}= \sqrt{\langle r_n^2 \rangle}-\sqrt{\langle r_p^2 \rangle}$,
were also calculated in the ground state and shown in Table \ref{table1}.
\begin{table}
\begin{center}
\begin{tabular}{|l|r|r|r|r|r|} \hline
asy-EoS       & $E_{sym}/A$    & L(MeV)  & $R_n$(fm) & $R_p$(fm) & $\Delta R_{np}(fm)$  \\ \hline
asysoft       &     29.9                 & 14.4    & 4.90  & 4.65    & 0.25 \\ \hline
asystiff      &     28.3                 & 72.6    & 4.95  &  4.65   & 0.30 \\ \hline
asysupstiff   &     28.3                 & 96.6    & 4.96  &  4.65   & 0.31 \\ \hline
\end{tabular}
\caption{The symmetry energy at saturation (in $MeV$), the slope parameters, neutron rms radius,
protons rms radius, neutron skin thickness for the three Asy-EOS.}
\label{table1}
\end{center}
\end{table}
The values obtained with our semi-classical approach
are in a reasonable agreement with those reported by employing other
models for similar interactions \cite{ref15}.
The neutron skin thickness is increasing with the slope parameter,
as expected from a faster reduction of the symmetry term on the surface 
\cite{ref3,baranPR}.
This feature has been discussed in detail in \cite{Colo}.

  To inquire on the collective properties of the pygmy dipole we excite the nuclear system
at the initial time $t=t_0 = 30fm/c$ by boosting along the $z$
direction all excess neutrons ($N_e = 32$) and in opposite direction all core
nucleons, while keeping the '{\rm c.m.}' of the nucleus at rest (Pygmy-like
initial conditions).
The excess neutrons were identified as the most
distant $N_e=32$ neutrons from the nucleus CM. Then the system is left to
evolve and the evolution of the collective coordinates $Y$, $X_c$ and $X$,
associated with the  different isovector dipole modes (pygmy, core and total dipole)
%introduced before,
is followed for $600 fm/c$
by solving numerically the equations (1).
%(\ref{vlasov}).
During the time evolution the number of t. p.
escaping from the system corresponds, on average, to less than half a neutron, 
while the total energy conservation is satisfied within $1.5\%$. 
% verified observing the same 
%quantities, in this case even better preserved. A similar conclusion regards the 
%density profiles for neutron and protons respectively.

%\begin{center}
%\MakeUppercase{{\bf RESULTS FOR DIPOLE OSCILLATIONS}}
%\end{center}
\subsection{Results for dipole oscillations}

\begin{figure}
\begin{center}
\includegraphics*[scale=0.36]{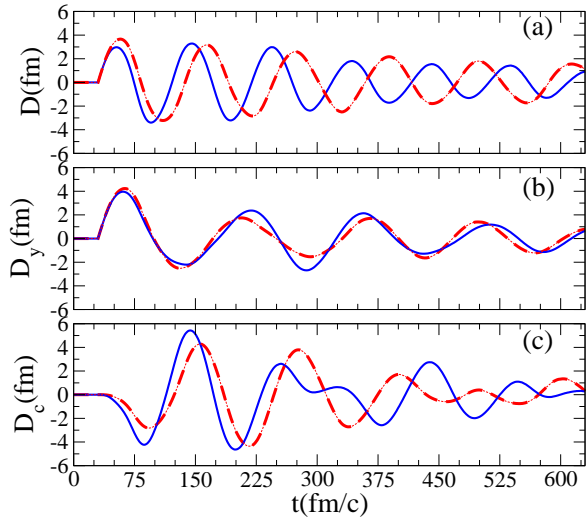}
\end{center}
\caption{(Color online) The time evolution of the total dipole $D$ (a),
of the dipole $D_y$ (b) and of core dipole $D_c$ (c),
for asysoft (the blue (solid) lines) and asysuperstiff (the red (dashed) lines) EOS. 
Pygmy-like initial excitation.}
\label{diptime}
\end{figure}

As shown in  figure \ref{diptime}, apart from the quite undamped oscillations of the $Y$ coordinate, 
we also remark that the core does not remain inert.
%In Fig. \ref{diptime} 
We plot the time evolution of the dipole
$D_y$, of the total dipole $D$ and core dipole $D_c$ moments, for two Asy-EoS.
As observed, 
while $D_y$ approaches its maximum
value, an oscillatory motion of the dipole $D_c$ initiates and
this response is symmetry energy dependent: the larger is the slope
parameter $L$, the more delayed is the isovector core reaction.
 This can be explained in terms of low-density (surface)
contributions to the vibration and therefore of the density behavior
of the symmetry energy below normal density: a larger L corresponds
to a larger neutron presence in the surface and so to a smaller
coupling to the core protons.
We see that the total dipole $D(t)$ is strongly affected by the presence of
isovector core oscillations,
mostly related to the isovector part of the effective interaction.
Indeed, $D(t)$ gets a higher oscillation frequency with respect to
$D_y$, sensitive to the Asy-EOS. The fastest
vibrations are observed in the asysoft case, which gives the largest
value of the symmetry energy below saturation. In correspondence the
frequency of the pygmy mode seems to be not much affected by the
trend of the symmetry energy below saturation, see also next figure
\ref{dipspectrum}, clearly showing the different nature,
isoscalar-like, of this oscillation. For each case we calculate the
power spectrum of $D_y$:
%\begin{equation}
$\displaystyle |D_y (\omega)| ^2 = |\int_{t_0}^{t_{max}} D_y(t) e^{-i\omega t} dt|^2$
%\end{equation}
and similarly for $D$. The results are shown in figure
\ref{dipspectrum}. The position of the centroid corresponding to
the GDR shifts toward larger values when we move from asysuperstiff (largest slope parameter $L$)
to asysoft EOS.
This evidences the importance of the volume, S-J component of the GDR in
 $^{132}Sn$.
The energy centroid associated with the PDR is situated below the GDR peak, at around $8.5 MeV$,
quite insensitive to the Asy-EOS,
pointing to an isoscalar-like nature of this mode.
 Hence the structure of the
dipole response can be explained in terms of the development of
isoscalar-like (PDR) and isovector-like (GDR) modes, as 
observed in asymmetric systems \cite{ref16}.
%\begin{figure}[h]
%\includegraphics[width=20pc]{fig6.eps}\hspace{2pc}%
%\begin{minipage}[b]{14pc}\caption{\label{imb_eloss}
%Imbalance ratios as a function of relative energy loss. 
%Upper panel: separately for 
%stiff (solid) and soft (dashed) Asy-EOS, and for 
%two parameterizations of the isoscalar part of the interaction: 
%MD 
%(circles and squares) and MI 
%(diamonds and triangles), in the projectile region (full symbols)
% and the target region 
%(open symbols).
%Lower panel: quadratic fit to all points for the stiff (solid), resp.
% soft (dashed) 
%Asy-EOS.}
%\end{minipage}
%\end{figure}  %\begin{figure}
\begin{figure}[h]
%\begin{center}
%\includegraphics*[scale=0.33]{pygmy_fou_i123_prc1.eps}
\includegraphics[width=18pc]{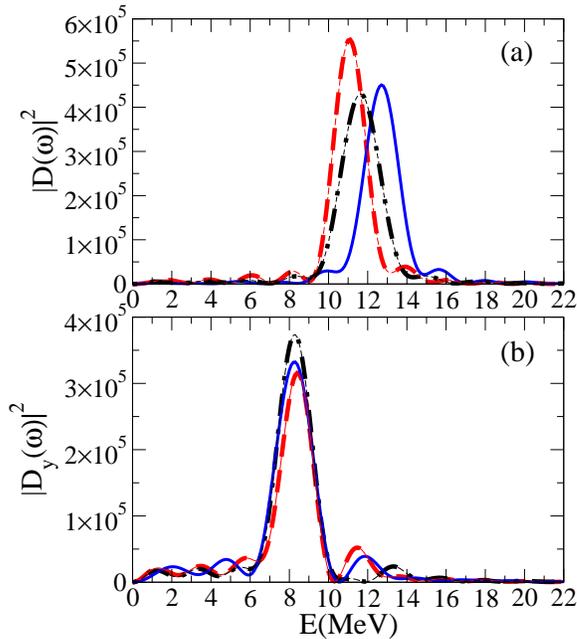}\hspace{2pc}
%\end{center}
\begin{minipage}[b]{14pc}
\caption{\label{dipspectrum}(Color online) The power spectrum of total dipole (a) and
of  the dipole $D_y$ (b) (in $fm^4/c^2$), for asysoft
(the blue (solid) lines), asystiff (the black (dot-dashed) lines) and asysuperstiff (the red (dashed) lines)
EOS. Pygmy-like initial conditions.
}
\end{minipage}
\end{figure}
Both modes are excited in the considered pygmy-like initial conditions. Looking
at the total dipole mode direction, that is close to the isovector-like normal mode,
one observes a quite large contribution in the GDR region. 
On the other hand, 
although the pygmy mode has a more complicated structure \cite{ref13},  
the $Y$ direction appears  closely related to it. 
Indeed a larger
response amplitude is detected in the pygmy region, see figure 2 (bottom).
%On the other hand, 
%considering the $Y$ direction, more closely related to the isoscalar-like mode, a larger
%response amplitude is detected in the pygmy region.

%%%%%%%%%%%%%%%qui
\begin{figure}
%\begin{center}
\includegraphics[width=18pc]{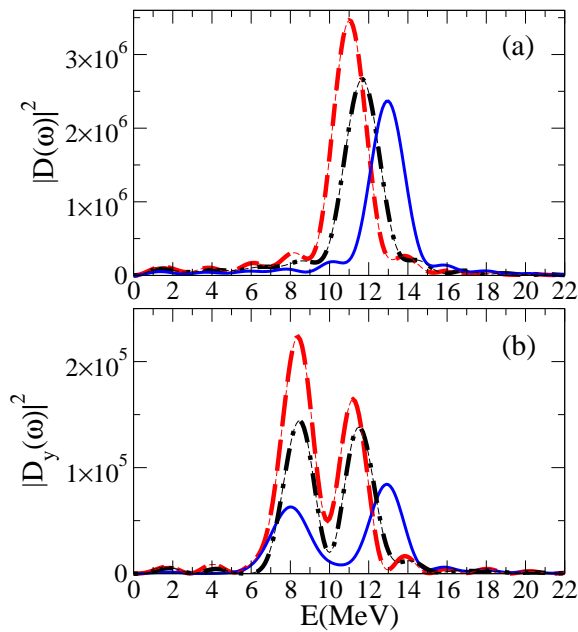}\hspace{2pc}
%\end{center}
\begin{minipage}[b]{14pc}
\caption{\label{gdrspectrum}(Color online) The same as in figure \ref{dipspectrum} but for a GDR-like
initial excitation.}
\end{minipage}

\end{figure}
To check the influence of the initial conditions on the dipole
response, let us consider the case of a  GDR-like excitation, corresponding
to  a boost of all neutrons against all protons, keeping the CM at rest.
%The initial collective energy correspond to first GDR excited state, around $15MeV$. 
Now the initial excitation favors the isovector-like mode and
even in the $Y$ direction we observe a sizable contribution in the
GDR region, see the Fourier spectrum of $D_y$ in figure \ref{gdrspectrum}. 
From this result it clearly emerges that 
a part of the $N_e$ excess neutrons is involved in a GDR type motion
and the relative weight depends on the symmetry
energy: more neutrons %, $N_y$, 
are involved in the pygmy mode in the 
asysuperstiff EOS case, in connection to the larger neutron skin size.
We have also checked that, if the coordinate Y is constructed
by taking the $N_y$ most distant neutrons (with $N_y < N_e$),
the relative weight increases in the PDR region. 
In any case, since part of the excess nucleons 
contributes to the GDR mode, a low EWSR value 
%than  the HOSM predictions corresponding to $N_y=N_e$ 
is expected in the PDR region. 
Indeed, in the Fourier power spectrum of $D$ in figure
\ref{gdrspectrum}, a weak response is seen at the pygmy frequency.
%%%%%%%%%%%%%%%%qui
In the case of the GDR-like initial excitation, i.e. boosting all neutrons against all protons, 
we can relate the strength function
to $Im(D(\omega))$ \cite{ref18} and then the corresponding
cross section can be calculated. 
%\footnote{We consider that a number of
%six oscillations in the presence of Landau damping
%will provide a reliable, though approximate, estimate of the 
%strength function.}.
Our estimate of the integrated cross section 
over the PDR region represents  $2.7 \%$
for asysoft, $4.4 \%$ for asystiff and $4.5 \%$ for asysuperstiff,
 out of the total cross section. Hence the EWSR
exhausted by the PDR is proportional to the skin thickness, in agreement
with the results of \cite{ref19}.
%From the total dipole acceleration, within a
%bremsstrahlung approach \cite{baran2001}, it is also possible to estimate
%the fraction of total photon emission probability in the PDR region
%out of the total dipole emission probability. We obtain a percentage
%of $4.7 \%$ for asysoft, $7.7 \%$ for asystiff and $9 \%$ for
%asysuperstiff EOS, consistent with the previous interpretation.

\section{Isospin equilibration and fragmentation mechanisms at Fermi energies}

In this energy range, reactions between charge asymmetric systems are charecterized by  
a direct isospin transport in binary events.
This process also involves the low density neck region and is sensitive to
the low density behavior of $E_{sym}$, see Refs.\cite{tsang92,isotr07} and references therein.
Moreover, it is now quite well established that the largest part of the reaction
cross section for dissipative collisions at Fermi energies goes
through the {\it Neck Fragmentation} channel, with intermediate mass
fragments (IMF) directly
produced in the interacting zone in semiperipheral collisions on short
time scales \cite{wcineck}. It is possible to predict interesting 
isospin transport effects also  for this 
fragmentation mechanism since clusters are formed still in a dilute
asymmetric matter but always in contact with the regions of the
projectile-like and target-like remnants almost at normal densities.  

Results on these mechanisms, obtained with the
SMF model, are discussed below. 
%Thus the low
%density behavior of the symmetry energy is concerned yet. 
%A neck of density below normal density develops between the two 
%heavy residues, the evolution of which is driven by the motion of the 
%spectators. 
%During this phase isospin is transferred to the neck due to the density 
%difference 
%between the neck and the residues; this effect is called isospin migration,
% which 
%leads to a more neutron-rich neck. In addition in collision systems with
% different asymmetry isospin is transported through neck due to the 
%asymmetry gradient
% leading to an equilibration of 
%the isospin of the residues, which has been called isospin diffusion. 
%Thus in asymmetric
% systems there is  a competion of isospin migration and diffusion.  

%In peripheral collisions 
%discussed here, 
%residues of about normal density are in contact with the neck region of 
%density below 
%saturation. At such low densities a stiff iso-EOS has a smaller 
%value but 
%a larger slope compared to a soft iso-EOS. Correspondingly we expect 
%opposite effects of 
%these models on the migration and diffusion of isospin.

\subsection{The isospin transport ratio }
In peripheral and semi-peripheral reactions, 
 it is interesting to look at the asymmetries of the various parts 
of the interacting system in the exit channel:
emitted particles,  projectile-like (PLF) 
and target-like fragments (TLF), and in  the  case of ternary (or higher
multiplicity) events,  
IMF's.
 In particular, one can  study  the
 so-called isospin transport ratio, which is defined as
%%% Hermann
\begin{equation}
%R^x_{P,T} = \frac{2x^M-(x^H+x^L)}{(x^H-x^L)} .
R^x_{P,T} = \frac{2(x^M-x^{eq})}{(x^H-x^L)}~,
\label{imb_rat}
\end{equation}
with $x^{eq}=\frac{1}{2}(x^H+x^L)$.
 Here, $x$ is an isospin sensitive quantity
that has to be investigated with respect to
equilibration.   We consider primarily the asymmetry 
$\beta= I = (N-Z)/A$,
but also other quantities, such as isoscaling coefficients, ratios of 
production of light
 fragments, etc, can be of interest \cite{WCI_betty}. %WCI  
The indices $H$ and $L$ refer to the symmetric reaction
between the
heavy  ($n$-rich) and the light ($n$-poor)  systems, while $M$ refers to the
mixed reaction.
$P,T$ denote the rapidity region, in which this quantity is measured, in
particular the
PLF and TLF rapidity regions. Clearly, this ratio is $\pm1$ in
the projectile
and target regions, respectively, for complete transparency, and oppositely
for complete
rebound, while it is zero for complete equilibration.
\begin{figure}[h]
\includegraphics[width=20pc]{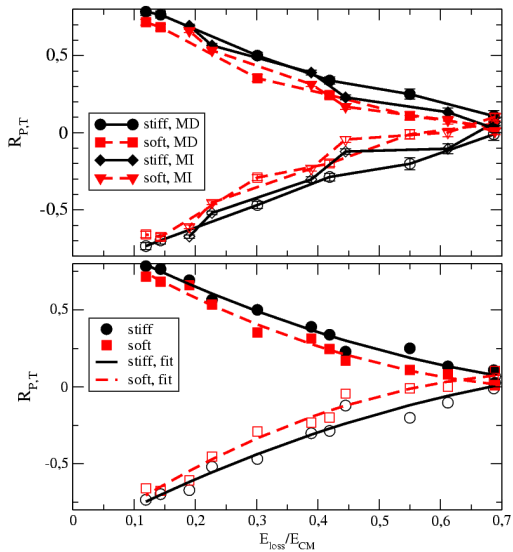}\hspace{2pc}%
\begin{minipage}[b]{14pc}\caption{\label{imb_eloss}
Isospin transport ratios as a function of relative energy loss. 
Upper panel: separately for 
stiff (solid) and soft (dashed) Asy-EOS, and for 
two parameterizations of the isoscalar part of the interaction: 
MD 
(circles and squares) and MI 
(diamonds and triangles), in the projectile region (full symbols)
 and the target region 
(open symbols).
Lower panel: quadratic fit to all points for the stiff (solid), resp.
 soft (dashed) 
Asy-EOS.}
\end{minipage}
\end{figure}  %\begin{figure}

%\begin{figure}[t]
%%\vskip 1.0cm
%\centering
%%\begin{picture}(0,0)
%%\put(135.0,0.0){\mbox{\includegraphics[angle=-90,width=7.0cm]{isotr11.ps}}}
%%\put{20.0,0.0}{mbox{\includegraphics[width=7.0cm]{isotr9.eps}}}
%%\end{picture}
%%\includegraphics[width=7.0cm]{erice3a.eps}
%%\hskip 0.5cm
%\includegraphics[width=7.5cm]{erice3b.eps}
%\caption{ 
%Left Panel.Imbalance ratios for $Sn + Sn$ collisions for 
%incident energies
%of 50 (left) 
%and 35 $AMeV$ (right) as a function of the impact parameter. Signatures of 
%the curves: 
%iso-EOS stiff (solid lines), soft (dashed lines); MD interaction (circles),
% MI interaction (squares); projectile rapidity ( full symbols, upper curves ),
% target rapidity ( open symbols, lower curves ).
%Right Panel. Imbalance ratios as a function of relative energy loss for both 
%beam energies. 
%Upper: Separately for 
%stiff (solid) and soft (dashed) iso-EOS, and for MD 
%(circles and squares) and MI 
%(diamonds and triangles) interactions, in the projectile region (full symbols)
% and the target region 
%(open symbols).
%Lower: Quadratic fit to all points for the stiff (solid), resp.
% soft (dashed) 
%iso-EOS.}
%\label{imb_eloss}
%\end{figure}

In a simple model one can show that the isospin transport ratio mainly depends on two
quantities: the strength of the symmetry energy and the interaction
time between the two reaction partners.
Let us take, for instance, the asymmetry $\beta$ of the PLF (or TLF) as the
quantity $x$.
%%% Hermann
At a first order approximation, in the mixed reaction this quantity relaxes
towards
its complete equilibration value, $\beta_{eq} = (\beta_H + \beta_L)/2$, as
\begin{equation}
\label{dif_new}
\beta^M_{P,T} = \beta^{eq} + (\beta^{H,L} -  \beta^{eq})~e^{-t/\tau},
\end{equation}
where $t$ is the time elapsed while the reaction partners stay in contact
(interaction time) and the damping $\tau$ is mainly connected to the strength 
of the symmetry energy \cite{isotr07}. 
Inserting this expression into Eq.(\ref{imb_rat}), one obtains
 %the following result for the imbalance ratio:
$ R^{\beta}_{P,T} = \pm e^{-t/\tau}$ for the PLF and TLF regions, respectively.
%From this simple result one sees that
%the imbalance ratio
%does not depend on the difference of asymmetries $(\beta_{H}- \beta_{L})$
%of the systems considered,
%at least at a first order level,
%which is due essentially to the normalization to the difference
%$(\beta_{H}- \beta_{L})$
%in the definition of $R$ in Eq.(\ref{imb_rat}).
Hence the isospin transport ratio can be considered as a good observable to 
trace back the strength
of the symmetry energy from the reaction dynamics
provided a suitable selection of the interaction time is performed.
%Our suggesiton is to look at the correlation with 
%the total kinetic energy loss of the dissipative collision, which 
%sets the
%time-scale of the process.
% We note, that the 
%imbalance ratio of eq.(\ref{imb_rat}) is constructed with the idea to 
%minimize the
%effects of  sequential decay \cite{tsang92}.
%[sentence deleted: "In ternary events we will, in
%addition to the asymmetries and the imbalance ratio, also discuss velocity
%correlations
% between the IMF and the residues, in order to gain information on the source
%of the IMF." This breaks the line of argments here and is discussed there.
%\subsubsection*{Correlation with kinetic energy loss}
The centrality dependence of the isospin ratio, for Sn + Sn collisions at 35 and 50 $MeV/u$,
has been investigated in experiments as well as in theory
\cite{tsang92,isotr07,yingxun},
and information about the stiffness of the symmetry energy has been extracted from
the analysis presented in \cite{yingxun}, based on the ImQMD model.  
%We report here a new analysis
% which appears experimentally more selective \cite{isotr07}. 

Here we investigate more in detail the relation between charge equilibration, 
interaction times and thermal equilibrium.  
Longer interaction times should be correlated to
a larger 
dissipation. It is then natural to look at the correlation between
the isospin transport ratio and the total kinetic energy loss.
In this way one can also better disentangle dynamical effects of the isoscalar 
and isovector part of the EOS, see \cite{isotr07}.

It is seen in figure \ref{imb_eloss} (top) that the curves for the 
asysoft EOS (dashed) are 
generally lower in the projectile region
 (and oppositely for the target region), i.e. show 
more equilibration, than those for the asystiff EOS, due to the higher value
of the symmetry energy at low density. 
To emphasize 
this trend, all  the values for the
stiff (circles) and 
the soft (squares) Asy-EOS, corresponding to different impact
parameters, beam energies and also to two possible parametrizations of the isoscalar part of the
nuclear interaction (with and without momentum dependence,
MD and MI), are collected together in the bottom part of the figure. 
One can see that all the points essentially follow a given line,
depending only on the symmetry energy parameterization adopted.  
% and fitted them by a quadratic curve. 
%It is seen that this fit 
%gives a good representation of the trend of the results.
%The difference between the curves for the stiff and soft iso-EOS in the 
%lower panel then isolates
% the influence of the iso-EOS from kinematical effects %depending on the 
%associated with the 
%interaction time. 
It is seen,
 that there is a systematic effect of the symmetry energy of the order 
of about 20 percent, 
which should be measurable. 
Moreover, we notice that the quantity $R$ is a rapidly decreasing function of the degree
of dissipation, $E_{loss}$,  reached in the collision. This can be explained 
in terms of 
dissipation mechanisms maily due to mean-field effects, as predicted by the
SMF model. Indeed, according to a mean-field picture, a significant degree of
thermal equilibrium 
(i.e. a considerable $E_{loss}$) would imply a rather long contact time between
the two reaction partners, thus certainly leading to isospin equilibration,
that needs a shorter time scale to be reached.   
The correlation suggested in figure \ref{imb_eloss}
should represent 
a general feature of isospin diffusion, as expected on the basis of  
dominant mean-field mechanisms,
and it would be of great 
interest to verify  it experimentally.

%As discussed above, this kind of analysis would help also in the
%comparison with the results of other theoretical models.

\vskip -1.0cm
\subsection{Isospin dynamics in neck fragmentation at Fermi energies}

\begin{figure}[h]
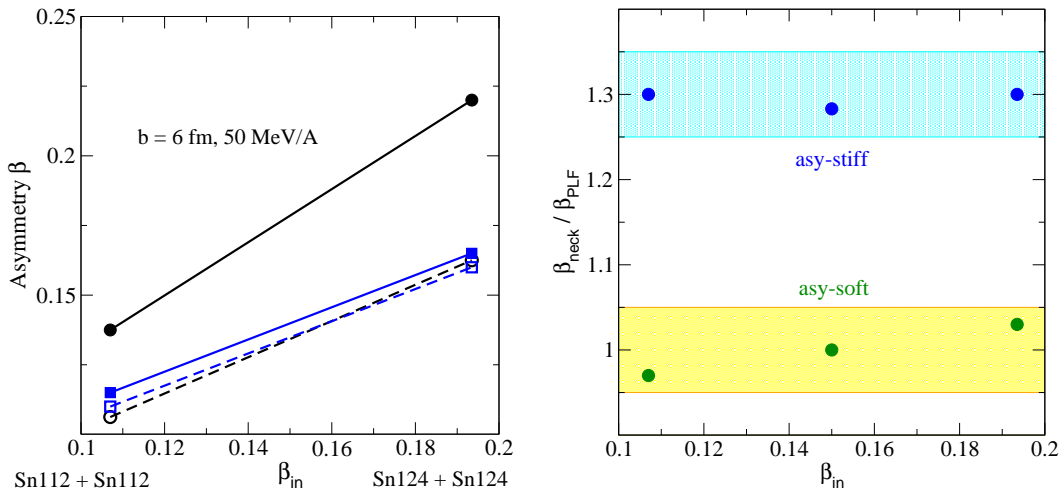

\centering
\vskip 1.0cm
 \includegraphics[width=16pc]{fig03_a.eps}%[scale=0.28]{erice5a.eps}
\hskip 0.5cm
 \includegraphics[width=16pc]{fig03_b.eps}%[scale=0.28]{name.eps}
%%\includegraphics[scale=0.30,angle=-90]{iwm3b.eps}
%\begin{picture}(0,0)
%\put(6.,140.){\mbox{\includegraphics[scale=0.30,angle=-90]{erice5b.eps}}}
%\end{picture}
%%\vskip -1.0cm
\caption{
Left panel: asymmetry of IMF's (circles) and PLF-TLF (squares), as a function of the
system initial asymmetry, for two Asy-EOS choices: asystiff (full lines) and
asysoft (dashed lines).  
Right panel: 
Ratio between the neck IMF and the PLF asymmetries, as a function of the system
initial asymmetry. The bands indicate the uncertainty in the calculations.}
%\vskip -2.0cm
\label{nzphi}
%\end{center}
\end{figure}

In presence of density gradients, as the ones occurring 
when a low-density neck region is formed between the two reaction
partners in semi-peripheral collisions,  the isospin transport
is mainly ruled by the density derivative of the symmetry energy  and so
we expect a larger neutron flow to
 the neck clusters for a stiffer symmetry energy around saturation  
\cite{baranPR}. This mechanism leads to the neutron enrichment of the
neck region (isospin migration). This is
shown in figure \ref{nzphi} (left), where the asymmetry of the neck and
PLF-TLF regions, obtained in neutron-rich reactions at 50 MeV/u, 
are plotted for two Asy-EOS choices. 
%The isospin dynamics can be directly extracted 
%from correlations between $N/Z$, $alignement$ and emission times of the $IMF$s.
%The alignment between $PLF-IMF$ and $PLF-TLF$ directions
%represents a very convincing evidence of the dynamical origin of the 
%mid-rapidity fragments produced on short time scales \cite{baranNPA730}. 
%The form of the
%$\Phi_{plane}$ distributions (centroid and width) can give a direct
%information on the fragmentation mechanism \cite{dynfiss05}. Recent 
%calculations confirm that the light fragments are emitted first, a general 
%feature expected for that rupture mechanism \cite{liontiPLB625}. 
%The same conclusion can be derived from {\it direct} emission time 
%measurements based on deviations from Viola systematics  observed
%in event-by-event velocity correlations between $IMF$s and the $PLF/TLF$ 
%residues
% \cite{baranNPA730,dynfiss05,velcorr04}. 
% We can figure out
%   a continuous transition from fast produced fragments via neck instabilities
%   to clusters formed in a dynamical fission of the projectile(target) 
%   residues up to the evaporated ones (statistical fission). Along this 
%   line it would be even possible to disentangle the effects of volume
%   and shape instabilities. 
%A neutron enrichment of the overlap ("neck") region is
%   expected, due to the neutron migration from higher (spectator) to 
%   lower (neck) density regions, directly related to
%%connected to 
%   the slope of the symmetry energy \cite{liontiPLB625}. 

From the experimental point of view, 
a new analysis has been recently published on Sn+Ni data at $35~MeV/u$
by the Chimera Collab.\cite{defilposter}.
% see figure \ref{nzphi} right panel.
A strong correlation between neutron enrichement and fragment alignement (when the 
short emission time selection is enforced) is seen, that points to 
%only with 
a stiff behavior of the symmetry energy ($L\approx 70~ MeV$), 
for which a large neutron
enrichment of neck fragments is seen (left). 
%, fig.\ref{nzphi} right panel \cite{isotr07}. 
%This represents an 
%evidence in favor of a relatively large slope ($L\approx 70~ MeV$) 
%around saturation. 
%We note a recent confirmation from structure data,
%i.e. from monopole resonances in Sn-isotopes \cite{garg_prl07}.   

%%%%%%%%%%%%%%%%%%%%%%%%%%%%%%%%%%%%%%%%%%%%%%%

%It is of interest to construct an observable, which quantifies these 
%effects and which does not 
%depend sensitively on measuring the absolute asymmetries, which are 
%changed by the secondary 
%evaporation. 
%%Considering that the asymmetries of the other reaction
%% products 
%%(residues and the gas) is not  equally  sensitive to the iso-EOS, it is attractive to look for ratios of
%%asymmetries.
%To this aim. 
In order to build observables less affected by secondary decay effects, 
in fig. 5 (right)  we consider the ratio of the asymmetries of the IMF's to those of the
residues ($\beta_{res}$) for stiff and soft Asy-EOS. 
%The results
% correspond to $b=6fm$ semiperipheral
%events, plotted  here  as a function of the initial isospin asymmetry of the
%colliding system.
%%For $b=8fm$ the behaviour is not very different,
%%except that the error bars are considerably larger. 
This quantity %between the asymmetry of IMF's and residues  
can be estimated
on the basis of simple energy balance considerations.
%In fact, isospin migration is due to the fact that the neck region
%has lower density with respect to the residues and the symmetry 
%energy is decreasing with density.
By imposing to get a  maximum  (negative) variation of 
$E_{sym}$ when transfering the neutron richness from
PLF and TLF towards the neck region, one obtains: 
%with respect to $\Delta\beta$ yields the ratio of asymmetries
\begin{equation}
\frac{\beta_{IMF}}{\beta_{res}} 
%= \frac{\beta+\Delta\beta}{\beta-\Delta\beta}
= \frac{E_{sym}(\rho_R)}{E_{sym}(\rho_I)} 
%= 1 +  \frac{E_{sym}(\rho_R)-E_{sym}(\rho_I)}{E_{sym}(\rho_I)}
\end{equation}
%\be
%\frac{\beta_{IMF}}{\beta_{Res}} \approx \frac{\beta + \Delta\beta}{\beta}
%%= \frac{\beta+\Delta\beta}{\beta-\Delta\beta}
%= 1 + \frac{E_{sym}(\rho_R)- E_{sym}(\rho_I)}{E_{sym}(\rho_R) + E_{sym}(\rho_I)}
%\ee
%The minimum of the variation $\delta E$ corresponds to:
%$$\delta\beta/\beta = \left( E_{sym}(\rho_R) - E_{sym}(\rho_{IMF})\right) /
%\left( E_{sym}(\rho_R) + E_{sym}(\rho_{IMF})\right)$$
%From this simple argument, one can see that the neutron enrichment of the
%neck region is proportional to the system asymmetry  $\beta$.
From this simple argument the ratio between the IMF and residue asymmetries should
depend only on symmetry energy properties and, in particular, on the difference of the 
symmetry energy corresponding to the residue and neck densities ($\rho_R$ and
$\rho_I$),
%between the residue and the neck regions, 
as appropriate 
for isospin migration.  
%symmetry energy properties and, in particular,
%on the ratio of the symmetry energies of the residue and the
%neck regions. 
It should also be larger than one, more so for the asystiff than
for the asysoft EOS.
%{\bf Check!!} 
It is seen indeed in figure 5 (right part), that this ratio %of IMF over residues asymmetry
is nicely dependent on the Asy-EOS only (being larger in the asystiff case) 
and not on the system considered.
If final asymmetries were affected in the same way by secondary evaporation 
in the case of neck and PLF fragments, then one could directly compare the
results of figure 5 (right) to data.  However, due to the different size and
temperature of the neck region with respect to PLF or TLF sources, 
de-excitation effects should be carefully checked with the help of
suitable decay codes. 

\vskip -1.0cm

\subsection{Comparison with the predictions of different transport codes}
A detailed investigation of isospin equilibration has been recently undertaken
within transport codes based on the molecular dynamics (QMD) approach \cite{yingxun}.
In comparison to the transport models considered before, mainly 
describing one-body effects (such as the SMF model, 
see Eq.(1)),
such approaches, where nucleons are represented as individual wave packets of fixed
compact shape (usually taken as gaussians), 
may lead to approximate descriptions of mean-field effects. 
%(surely leads to some approximation in the description...) 
%one expects that the description of mean-field
%effects is approximated, 
On the other hand, fluctuations and correlations should be well
reproduced, especially in
the exit channel of multifragmentation events.
As shown in Ref.\cite{yingxun}, where charge equilibration is investigated for $Sn + Sn$ reactions
at 35 and 50 $MeV/u$, 
the ImQMD code predicts a quite different behavior with respect to SMF:
the isospin transport ratio exhibits a rather flat behavior as a function
of the impact parameter.  
This seems to indicate that, even in the case of central collisions, the contact time
between the two reaction partners remains rather short, the dissipation mechanisms being mostly due
to many-body correlations rather than to mean-field effects. Thus the more explosive
dynamics would lead to the lower degree of isospin equilibration observed.

To examine more in detail the origin of the observed discrepancies, 
results concerning IMF ($Z>2$) properties, obtained with the SMF and ImQMD codes, are compared in figure 6.
In the left panel, the average total charge per event, associated with IMF's, is plotted
as a function of the reduced rapidity, for the reaction $^{124}$Sn + $^{124}$Sn at 50 MeV/u and 
impact parameters b = 6 and 8 fm. 
From this comparison it is clear that in ImQMD a larger number of light IMF's, distributed
over all rapidity range between PLF and TLF, are produced.
On the other hand, mostly binary or ternary events are observed in SMF, with light IMF's
located very close to mid-rapidity. Then the different reaction dynamics predicted by the two codes
may explain the different results seen for isospin equilibration especially
in semi-peripheral and central reactions (b $\approx$ 4-6 fm).
The fast ImQMD fragmentation dynamics inhibits nucleon exchange and charge equilibration. 
%leads to less charge equilibration, while 
On the other hand, in SMF dissipation
is dominated by mean-field mechanisms, acting over longer time intervals and leading to stronger
equilibration effects.
\begin{figure}
\begin{center}
\includegraphics[width=30pc]{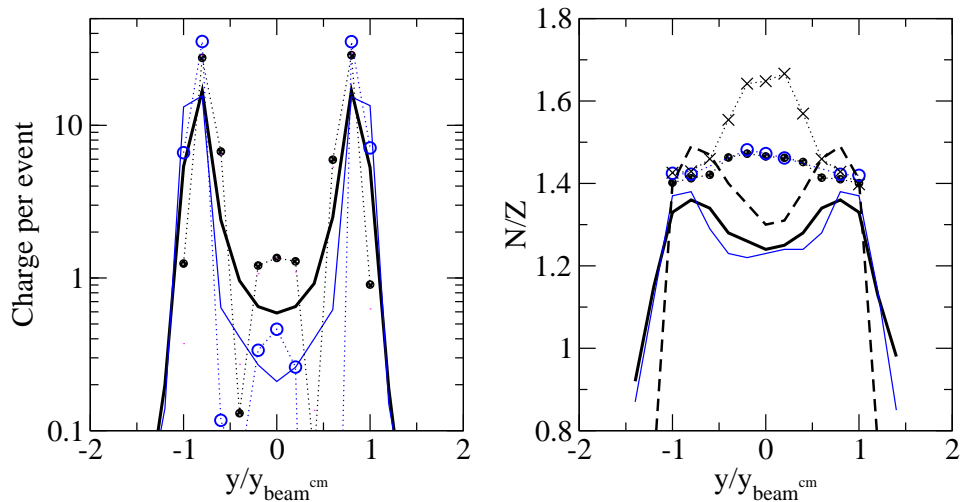}%\hspace{4pc}%
\end{center}
%\begin{minipage}[b]{14pc}
\caption{\label{confro}
Let panel: Average total charge per event, associated with IMF's, as a function of the reduced 
rapidity, obtained in the reaction $^{124}$Sn + $^{124}$Sn at 50 MeV/u.
Results are shown for ImQMD calculations at b = 6 fm (thick line) and b = 8 fm (thin line)
and for SMF calculations at b = 6 fm (full circles) and b = 8 fm (open circles). 
A soft interaction is considered for the symmetry energy.
Right panel: N/Z of IMF's as a function of the reduced rapidity. Lines and symbols are
like in the left panel. 
Results corresponding to a stiff asy-EOS are also shown for ImQMD (dashed line) and
SMF (crosses), for b=6 fm.}
%\end{minipage}
\end{figure}  %\begin{figure} 
Results on the neutron content of the neck region are illustrated in the right panel of figure 6,
that shows the global N/Z of IMF's as a function of the reduced rapidity. As discussed above,
SMF calculations clearly predict a larger N/Z for IMF's produced at mid-rapidity,  
with respect to PLF and TLF regions (isospin migration effect). 
The effect is particularly pronounced in the case of the asystiff parametrization. 
On the contrary, ImQMD calculations predict a minimum of the N/Z ratio at mid-rapidity.
The reasons of the these differences  need to be further investigated.

%This indicates a different nature of the dissipation
%mainly due to two-body mechanism.
%....The neutron enrichment of the neck region is not observed ...
%It would be important to check the reaction dynamics before concluding
%on the extraction of the symmetry energy.  

\vskip -1.0cm

\section{Conclusions}

We have reviewed some aspects of the rich phenomenology associated with nuclear 
reactions, from which interesting hints
are emerging to constrain the nuclear EOS and, in particular, the largely 
debated density behavior of the symmetry energy.
 Information on the low density region 
can be accessed in reactions from low to intermediate energies, 
where collective excitations and fragmentation mechanisms are dominant.
We have shown, within  a
semi-classical Landau-Vlasov approach, the existence, in neutron
rich nuclei, of a collective pygmy dipole mode determined by the
oscillations of some excess neutrons against the nuclear
core. From the transport simulations the PDR energy centroid
for $^{132}Sn$ appears around $8.5$ $MeV$, rather insensitive to the
density dependence of the symmetry energy and well below the GDR peak.
This supports the isoscalar-like character of this collective
motion. A complex pattern, involving the coupling  of the neutron skin with
the core dipole mode, is noticed.
% The transport model
%indicates that part of the excess neutrons $N_e$ are coupled to the 
%GDR mode and gives some hints about 
%the number of neutrons, $N_y$, actually participating in the pygmy mode.
This reduces considerably the EWSR acquired by the PDR, our numerical
estimate providing values well below $10 \%$, but proportional to the
symmetry energy slope parameter $L$, that affects the number of
excess neutrons on the nuclear surface. 
%We consider these effects as
%related also to the S-J component of the dipole dynamics
%in medium-heavy nuclei. It is therefore interesting to extend the
%present analysis to lighter nuclei, like Ni or Ca isotopes, where
%the Goldhaber-Teller component can be more important. 

%We would like to mention that such self-consistent, transport  approaches,
%can be valuable in exploring the collective response of other
%mesoscopic systems where similar normal modes may manifest,
%see [20].

Concerning nuclear reactions at Fermi energies, 
we have discussed some results on isospin sensitive observables.
%aspects of the phenomenology associated with nuclear 
%reactions at Fermi energies, 
%from which hints
%are emerging to constrain the EOS of asymmetric matter below normal density.
In particular, we have concentrated our analysis on the charge equilibration
mechanism (and its relation to energy dissipation) and on the neutron-enrichment of the neck region in semi-peripheral reactions.  From the study of the latter mechanism,
for which new experimental evidences have recently appeared \cite{defilposter}, hints are
emerging towards a stiff behavior of the symmetry energy around normal 
density.    
This is compatible with recent results from structure data,
see for instance the review article \cite{structure_data}.

Finally, we have also considered results of different transport codes (see section 4.3),
undertaking a comparison of ImQMD and SMF predictions for selected observables. 
Due to the different role of mean-field effects, vs. fluctuations
and many-body correlations, in the two codes, the description of the 
reaction dynamics is
model dependent (see also \cite{coup}). 
This affects also isospin sensitive observables. 
Thus the whole reaction path  needs to be carefully checked against experimental data
in order to get decisive conclusions about the density dependence of the symmetry energy.      
%and, in particular, the largely 
%debated density behavior of the symmetry energy.
%The greatest theoretical uncertainties concerns the high density 
%domain, 
%%behavior of
%%the symmetry energy, i.e. the domain 
%that has the largest impact on the
%understanding of the properties of neutron stars. 
%%This regime can be explored in terrestrial laboratories by using relativistic
%%heavy ion collisions of charge asymmetric nuclei. Differential 
%%collective flows,
%%particle and meson production are promising observables.  
%On the other hand, the behavior of the symmetry energy at low density 
%can be accessed in reactions from low to intermediate energies, 
%where collective excitations and fragmentation mechanisms are dominant.

\vskip 1.0cm
\noindent
{\bf Acknowledgements}

\noindent
We warmly thank H.Wolter and M.Zielinska-Pfabe for inspiring discussions.

\noindent 
This work for V. Baran was supported by a grant of the Romanian National
Authority for Scientific Research, CNCS - UEFISCDI, project number PN-II-ID-PCE-2011-3-0972.
For B. Frecus this work was supported by the strategic grant POSDRU/88/1.5/S/56668.

\vskip 1.0cm
\noindent
%{\bf References}
%\section*{References}
%\vskip 0.5cm
%We warmly thank A.Drago and A.Lavagno for the  
%collaboration on the
%mixed hadron-quark phase transition at high baryon and isospin density.

%One of authors, V. B. thanks for warm hospitality at Laboratori
%Nazionali del Sud, INFN. This work was supported in part by the Romanian
%Ministery for Education and Research under the contracts PNII, No.
%ID-946/2007.

                                    %
%%%%%%%%%%%%%%%%%%%%%%%%%%%%%%%%%%%%%%%%%%%%%%%%%%%%%%%%%%%%%%%%%%%%%%%%%

%\include{nn06_ditorobib}

%%%%%%%%%%%%%%%%%%%%%%%%%%%%%%%%%%%%%%%%%%%%%%%%%%%%%%%%%%%%%%


\begin{thebibliography}{99}

%%%%%%%%%%%%%%%%%%%%%%%%%%%%%%%%%%%%%%%%%%%%%%%%%%%%%%%%%%%%%%%%%%%%%
\bibitem{cons} Danielewicz P, Lacey R and Lynch WG 2002 {\it Science} 
{\bf 298} 1592

\bibitem{Lattimer}
Lattimer JM and Prakash M 2007 {\it Phys.Rep} {\bf 442}

\bibitem {Colo}
Carbone A {\it et al} 2010 
%Andrea; Colo Gianluca; Bracco Angela; et al.Source: 
{\it Phys. Rev.} {\bf C81} 041301   %DOI:

\bibitem{Isospin01} {\it Isospin Physics in Heavy-ion Collisions at
Intermediate Energies}, Eds. Li BA and Schr\"oder WU, Nova Science Publishers (2001, New York)

\bibitem{baranPR} Baran V, Colonna M, Greco V and Di Toro M 2005 
{\it Phys. Rep.} {\bf 410} 335

\bibitem{WCI_betty}
Colonna M and Tsang MB 2006 
{\it Eur. Phys. J} A {\bf 30} 165,
and references therein

\bibitem{baoPR08} 
Li BA, Chen LW and Ko CM 2008 
{\it Phys. Rep.} {\bf 465} 113

\bibitem{ref1}
Paar N, Vretenar D, Khan E, Col\'o G 2007
{\it Rep. Prog. Phys} {\bf 70} 691


\bibitem{chomazPR}
Chomaz P, Colonna M and Randrup J 2004
{\it Phys. Rep.} {\bf 389} 263

\bibitem{SMF}
Colonna M {\it et al} 1998 {\it Nucl. Phys.} {\bf A642} 449;
Rizzo J, Chomaz Ph, Colonna M 2008
{\it Nucl. Phys.} {\bf A806} 40  and references therein 

\bibitem{ref2}
Klimkiewicz A {\it et al} 2007
{\it Phys. Rev.} {\bf C76} 051603(R)

\bibitem{ref3}
Yoshida S, Sagawa H 2004
{\it Phys. Rev.} {\bf C69} 024318;
2006 {\it Phys. Rev.} {\bf C73} 044320

\bibitem{ref4}
Piekarewicz J 2006
{\it Phys. Rev.} {\bf C73} 044325

%\bibitem{ref5}
%Carbone A {\it et al} 2010
%{\it Phys. Rev.} {\bf C81} 041301(R)

\bibitem{ref6}
 Savran D {\it et al} 2008
{\it Phys. Rev. Lett.} {\bf 100} 232501

\bibitem{ref7}
Wieland O {\it et al} 2010
{\it Phys. Rev. Lett.} {\bf 102} 092502;
Wieland O, Bracco A 2011
{\it Prog.Part. Nucl. Phys}. {\bf 66} 304;
Toft HK {\it et al} 2010
{\it Phys. Rev.} {\bf C81} 064311

\bibitem{ref8}
Tonchev AP {\it et al} 2010
{\it Phys. Rev. Lett.} {\bf 104} 072501

\bibitem{ref9}
Makinaga A {\it et al} 2010
{\it Phys. Rev.} {\bf C82} 024314

\bibitem{ref10}
 Steinwedel H, Jensen JHD 1950
{\it Z. Naturf.} {\bf 5A} 413

\bibitem{ref11}
 Baran V, Rizzo C, Colonna M, Di Toro M, Pierroutsakou D 2009
{\it Phys. Rev.} {\bf C79} 021603(R)

\bibitem{ref12}
Roca-Maza X, Pozzi G, Brenna M, Mizuyama K, Col\'o G 2012
{\it Phys. Rev.} {\bf C85} 024601

\bibitem{ref13}
Urban M 2012 {\it Phys. Rev.} {\bf C85} 034322

\bibitem{ref17}
Baran V, Frecus B, Colonna M and Di Toro M 2012 
{\it Phys. Rev.} {\bf C85} 051601(R)

%\bibitem{ref14}
% Baran V, Colonna M, Di Toro M, Greco V 2005
%{\it Phys. Rep.} {\bf 410} 335

\bibitem{ref15}
Paar N, Nicsik T, Vretenar D, Ring P 2005
{\it Phys. Lett.} {\bf B606} 288

\bibitem{ref16}
Baran V, Colonna M, Di Toro M, Greco V 2001
{\it Phys. Rev. Lett}. {\bf 86} 4492; Colonna M, Chomaz P, Ayik S 2002
{\it Phys. Rev. Lett.} {\bf 88} 122701



\bibitem{ref18}
Calvayrac F, Reinhard PG, Suraud E 1997
{\it Ann. Phys.} {\bf 225} 125 

\bibitem{ref19}
Inakura T, Nakatsukasa T, Yabana K 2011
{\it Phys. Rev.} {\bf C84} 021302(R)





%\bibitem{baldoreview} Baldo M {\it et al} 2004 {\it Nucl. Phys.} A {\bf 736} 241
  

\bibitem{tsang92} 
Tsang MB {\it et al} 2004 {\it Phys. Rev. Lett.} 
 {\bf 92} 062701 


\bibitem{isotr07} Rizzo J {\it et al} 2008  
{\it Nucl.Phys.}  A {\bf 806} 79

\bibitem{yingxun}
Tsang MB {\it et al} 2009 %; Zhang, Yingxun; Danielewicz, P.; et al.
{\it Phys. Rev. Lett.} {\bf 102}  122701;
%Title: Isospin diffusion and equilibration for Sn plus Sn collisions at E/A=35 MeV
Sun ZY {\it  et al} 2010
{\it Phys. Rev.} {\bf C82}  051603
%DOI: 10.1103/PhysRevC.82.051603   Published: NOV 23 2010
%Times Cited: 8 (from Web of Science) 
  
%DOI: 10.1103/PhysRevLett.102.122701   Published: MAR 27 2009
%Times Cited: 143 (from Web of Science)

\bibitem{wcineck}
Di Toro M, Olmi A and Roy R 2006 {\it Eur. Phys. Jour.}
 {\bf A30} 65 

%\bibitem{isotr05} 
%Baran V {\it et al} 2005 {\it Phys.Rev.} C {\bf 72} 064620

\bibitem{tsang_exp}
Tsang MB {\it et al} 2011
{\it Prog. Part. Nucl. Phys.} {\bf 66}   400
%-404   DOI: 10.1016/j.ppnp.2011.01.041   Published: APR 2011



\bibitem{defilposter}
De Filippo E {\it et al} (Chimera Collab.) 2012
 {\it Phys. Rev.} {\bf C86} 014610
%{\it Time scales and isospin effects on reaction dynamics},
% NN06 Conf., Rio de Janeiro, August 2006, and {\it Isospin signals in
%reaction dynamics}, Int.Conf.on Nuclear Fragmentation, Antalya 2007




\bibitem{structure_data}
%Li T, Garg U {\it et al} 2007
%{\it Phys. Rev. Lett.} {\bf 99} 162503
Tsang MB {et al} 2012 arXiv:1204.0466

\bibitem{coup} 
Coupland D {\it et al} 2011
{\it Phys. Rev.} {\bf C84}  054603   
%DOI: 10.1103/PhysRevC.84.054603   Published: NOV 3 2011
%Times Cited: 4 (from Web of Science)





%\bibitem{BaoINTJE7} B.A. Li, C.M .Ko, W. Bauer,
%Int. J. Mod. Phys. {\bf E7} (1998) 147 












%\bibitem{theo04}
%Gaitanos T et al. 2004
%{\it Nucl. Phys.} A {\bf 732} 24








%\bibitem{FlibPRL77}
%S. Flibotte et al.,
%{\em Phys. Rev. Lett.} {\bf 77} (1996) 1448.

%\bibitem{CinNC111}
%M. Cinausero et al.,
%{\em Nuovo Cimento} {\bf 111} (1998) 613.

%\bibitem{PierrouEPJA16}
%D. Pierroutsakou et al.,
%{\em Eur. Phys. Jour.} {\bf A16} (2003) 423,
% {\em Nucl. Phys.} {\bf A687} (2003) 245c.

%\bibitem{AmoPRC29}
%F.Amorini et al.,
%{\em Phys. Rev.} {\bf C69} (2004) 014608.



%\bibitem{BaranNPA600}
%V. Baran et al.,
%{\em Nucl. Phys.} \textbf{A600} (1996) 111.

%\bibitem{BaranNPA679}
%V. Baran et al.,
%{\em Nucl. Phys.} \textbf{A679} (2001) 373.

%\bibitem{SimenPRL86}
%C. Simenel et al.,
%{\em  Phys. Rev. Lett.} \textbf{86} (2000) 2971.



%\bibitem{PierrouLNS}
%D.Pierroutsakou et al.,
%LNS  exp. proposal  2006, PAC approved.

%\bibitem{luca04}
%L. Bonanno,
%{\em Effetti di radiazione diretta dipolare sulla sintesi degli elementi
% superpesanti}, Master Thesis, Catania Univ. 2004.











%\bibitem{shiPRC68}
%L. Shi, P. Danielewicz, Phys. Rev.{\bf C68} (2003) 064604 







%\bibitem{wolter_iwm}
%H.H.Wolter et al., 
%{\em Isospin Sensitive Observables in Peripheral Collisions},
% IWM Int.Workshop,
%SIF, Conf.Proc.Vol. 95 (2008) p. 287-294..








% {\em Phys.Rev.Lett.} submitted



%\bibitem{dissip}
%It is interesting to note that opposite effect of the symmetry stiffness on 
%dissipation is expected at the Fermi energies, i.e. larger interaction times 
%for the asysoft case, see \cite{colonnaPRC57}. This can be easily understood 
%since at higher energies we are testing suprasaturation densities in the 
%interacting region, with opposite trend of the symmetry repulsion.

%\bibitem{colonnaPRC57}
%M.Colonna, M.Di Toro, G.Fabbri, S.Maccarrone, 
%{\em Phys. Rev.} {\bf C57} (1998) 1410.


%\bibitem{simenelPRL86}
%C.Simenel, P.Chomaz, G.de France,
%{\em Phys. Rev. Lett.} {\bf 86} (2000) 2971.

%\bibitem{baranPRL87}
%V.Baran, D.M.Brink, M.Colonna, M.Di Toro,
%{\em Phys. Rev. Lett.} {\bf 87} (2001) 335.

%\bibitem{pierrouPRC71}
%D.Pierroutsakou et al., 
%{\em Phys. Rev.} {\bf C71} (2005) 054605.



%\bibitem{paganoNPA734}
%A.Pagano et al. (Chimera Collab.),
%{\em Nucl. Phys.} {\bf A734} (2004) 504c.



%\bibitem{wilcz}
%It has been proposed to call such Viola-violation-correlation plot as 
%$Wilczynski-2~Plot$. Apart the origin from discussions with Janusz at 
%the LNS/INFN Catania, in fact this correlation represents also a 
%$chronometer$ of the fragment formation mechanism. In this sense it is a 
%nice Fermi energy complement of the famous $Wilczynski-Plot$ which gives 
%the time-scales in Deep-Inelastic Collisions.

%\bibitem{milazzo}
%P.M.Milazzo et al. (Multics Collab.),
%{\em Phys. Lett.} {\bf B509} (2001) 204; 
%{\em Nucl. Phys.} {\bf A703} (2002) 466.

%\bibitem{colinPRC67}
%J.Colin et al.,
%{\em Phys. Rev.} {\bf C67} (2003) 064603.







%\bibitem{ZuoPRC72} W.Zuo, L.G.Cao, B.-A.Li, U.Lombardo, C.W.Shen,
% {\em Phys. Rev.} {\bf C72} (2005) 014005.

%\bibitem{DalenarPRL95} E. van Dalen, C.Fuchs, A.F\"assler,
%{\em Phys. Rev. Lett.} {\bf 95} (2005) 022302.

%\bibitem{OlliPRD46} 
%J.Y. Ollitrault, 
%{\em Phys. Rev.} {\bf D46} (1992) 229.



%\bibitem{BaoPRL82} 
%B.A. Li and A.T. Sustich, 
%{\em Phys. Rev. Lett.} {\bf 82} (1999) 5004.

%\bibitem{fopi_v1} A.Andronic et al., FOPI Collab., 
%{\em Phys. Rev.} {\bf C67} (2003) 034907.

%\bibitem{fopi_v2} A.Andronic et al., FOPI Collab., 
%{\em Phys. Lett.} {\bf B612} (2005) 173.

%\bibitem{GalePRC41} Gale C, Bertsch GF, Das Gupta S 1990 
% {\it Phys.Rev.} C {\bf 41} 1545

%\bibitem{BombaciNPA583} Bombaci I et al 1995
%{\it Nucl.Phys.} A {\bf 583} 623







%\bibitem{Klahn06}
% T.Kl\"ahn et al.,
%{\em Phys. Rev.} {\bf C74} (2006) 035802.

%\bibitem{Liubo07}
% B.Liu et al., 
%{\em Phys. Rev.} {\bf C75} (2007) 048801.




%\bibitem{NJL}
%Nambu Y, Jona-Lasinio G 1961
%{\it Phys. Rev.} {\bf 122} 345; {\bf 124} 246


%\bibitem{ScalonePLB461}
%L.Scalone, M.Colonna, M.Di Toro, 
%{\em Phys. Lett.} {\bf  B461} (1999) 9.


%%%%%%%%%%%%%%%%%%%%%%%%%%%%%%%%%%%%%%%%%%%%%%%%%%%%%%%%%%%%%%
\end{thebibliography}
\end{document}